\documentclass[11pt]{article}
\usepackage{graphicx}       
\newcommand{\re}{\mathrm{e}}
\newcommand{\hp}{\mathrm{h}}
\newcommand{\ri}{\mathrm{i}}

\newcommand{\cp}{\mathcal P}
\newcommand{\cs}{\mathcal S}
\RequirePackage{amssymb}
\usepackage{geometry} 
\begin{document}

\begin{center}
{\bf Coherent Distributions and Quantization} \\[2cm] 
Marius Grigorescu \\[3cm]  
\end{center}
\noindent
$\underline{~~~~~~~~~~~~~~~~~~~~~~~~~~~~~~~~~~~~~~~~~~~~~~~~~~~~~~~~
~~~~~~~~~~~~~~~~~~~~~~~~~~~~~~~~~~~~~~~~~~~~~~~~~~~~~~~~~~}$ \\[.3cm]
This work\footnote{Based on the research proposal 338620ERC-2013-ADG
"Functional Coherent Distributions on Granular Phase-Space and Quantization", 
November 22 (2012).} presents a selective review of results concerning the mathematical interface 
between the classical and quantum aspects encountered in problems such as the 
nuclear mean-field dynamics or quantum Brownian motion. It is shown that the 
main difference between classical and quantum behavior arises from the 
coherence properties of the phase-space distributions known as "action waves" 
and Wigner functions.  The quantum wave functions appear as elementary degrees
of freedom for the phase space granularity.   \\
$\underline{~~~~~~~~~~~~~~~~~~~~~~~~~~~~~~~~~~~~~~~~~~~~~~~~~~~~~~~~~~~~~
~~~~~~~~~~~~~~~~~~~~~~~~~~~~~~~~~~~~~~~~~~~~~~~~~~~~~}$ \\

\newpage

\section{Introduction} 
In classical statistical mechanics, the entropy of  a many-body system  is defined with 
respect to a partition of the  $2n$-dimensional one-particle phase-space in elementary 
cells, but the size of these cells is not specified.   The occurrence  of $\hp =6.626 
\times 10^{-34}$ J$\cdot$s  in the Planck distribution  for  thermal radiation was 
considered as evidence for such a granular structure, with cells ("quantum states") of 
volume $\hp^n$.  The old quantum mechanics maintained this view, as for the integrable 
systems in action-angle coordinates a cell structure was introduced by the set of stationary orbits (invariant tori) 
selected by the Bohr-Sommerfeld  quantization conditions. In  quantum mechanics however,  
the  physical states of microparticles are described by rays in an abstract Hilbert 
space,  and its main elements (wave-functions and operators)  are expressed  in terms of 
only  $n$  coordinates.  This formalism,  though quite successful  in  atomic physics, 
is not complete,  because   it requires additional rules  for "quantization" (e.g.  
algebraic (Dirac), geometric \cite{am, bk, nw, gs}) and interpretation (e.g. probabilistic 
interpretation of the scalar product).  These rules provide an interface between the 
physical observables defined at classical level and the abstract Hilbert space  in some 
important situations of interest  (e.g. the measurement process), but are not enough 
general to join classical and quantum mechanics into a single theory.  One of the 
difficulties in formulating such a theory is that while thermal fluctuations in classical 
systems can be explained by the action of random forces, the origin of the quantum 
fluctuations is unknown.  The conceptual gap between classical and quantum mechanics is 
illustrated  for instance by the early (1927) Bohr-Einstein debate on the intrinsic 
statistical character of quantum mechanics,  the paradox of the "Schr\"odinger's cat 
experiment", or the Zeno paradox at the continuous measurement process \cite{zp}. There 
was also a constant effort to bridge this gap by non-standard theories such as the 
thermodynamics of the isolated particle of L. de Broglie,  the Bohmian interpretation of 
quantum mechanics, or the hypothesis of spontaneous  wave  function collapse   
\cite{grw}.  \\ \indent
Considering  quantum mechanics as fundamental, in the case of a mixed system containing 
classical and quantum components  one can  start with a quantum description of the whole 
system, and then perform  partial tracing over the variables known as classical.  Along 
this line, the classical dynamics of a nuclear collective model can be derived  as 
constrained quantum dynamics if in the quantum time-dependent variational principle 
(TDVP) for the Schr\"odinger Lagrangian  $L_\psi = \langle \psi \vert \ri \hbar \partial_t - 
\hat{H} \vert \psi \rangle $,  $\psi$ belongs to a  suitable manifold of trial wave 
functions  (e.g. coherent states) \cite{dt, cev}.  A possible mechanism to generate such a 
collective phase-space  as a symplectic submanifold of the quantum Hilbert space  is the  
spontaneous symmetry breaking \cite{cpss}.  \\ \indent
 Classical  degrees of freedom in  nuclear, atomic or molecular systems can be introduced  
using  the  Born-Oppenheimer approximation. However, attempts to formulate a theory of  
genuine mixed classical-quantum dynamics appear in 1994, in the form of coupled 
Hamilton-Schr\"odinger  \cite{gc}  or  Hamilton-Heisenberg   \cite{aa}  
equations,  derived from a mixed  TDVP. The mixed Lagrangian contains a classical part 
$L_{cl}(x, \dot{x}) $,  the quantum part $L_\psi$, and an interaction term $L_{int}(x, 
\psi)$ depending both on the coordinates  $x(t)$  of the classical component and the 
quantum wave function $\psi(q,t)$.  \\ \indent
In the case of a quantum particle coupled to a thermal bath of classical harmonic 
oscillators by a bilinear interaction the Hamilton-Schr\"odinger  equations can be reduced 
to  a  non-linear Schr\"odinger-Langevin equation,  containing noise and  non-linear 
friction terms  (nonlinearity due to the  backreaction of the classical environment) \cite{gc}. 
Further applications present two variants of this approach:
 
-  for a many-body quantum system,  described within some mean-field approximation,  $\psi$ 
 is constrained to a symplectic submanifold  of the quantum Hilbert space.  The example of 
a superfluid quantum many-fermion system in a thermal environment was  considered  in 
\cite{tdhfb1, tdhfb2}. 

- for the study of decoherence  $L_\psi$  can be  expressed  in terms of the density matrix, 
such that instead of the Schr\"odinger-Langevin equation  one obtains  a quantum 
Liouville-Langevin  equation  \cite{ddq}. Applications to atomic  transition rates in  
thermal radiation  field and decoherence time for a two-level system with  Ohmic dissipation 
are presented in \cite{ddq}. Numerical integration shows that dissipation produced by the 
non-linear friction term alone (at zero temperature) resembles the spontaneous decay 
obtained when the classical environment is quantized. The realistic  situation  of  dissipative  
atomic tunneling in an asymmetric quartic (double well) potential at finite 
temperature  is presented in \cite{haat}. Analytically it was shown that if the nonlinear 
friction term is neglected, then by taking the ensemble average, the results are  equivalent 
to the integration of a quantum  Fokker-Planck equation for the density matrix \cite{haat}. 

The extensive literature on these subjects also includes a detailed analytical 
study of the environmental decoherence during  macroscopic quantum tunneling  in a cubic 
potential  using a quantum Kramers equation for the reduced Wigner function of the 
tunneling particle   \cite{cv}, or a variational principle describing  a classical 
statistical ensemble on the configuration space interacting with a quantum system, applied 
to  couple  quantum matter fields and classical metric  \cite{icqe}.
 
The variants  \cite{gc, tdhfb1, ddq}  are summarized  in  \cite{vpr}, using  for  the 
Lagrangian  $L_{tot}$  a more compact form, with the term $L_\psi$ alone, but the  trial 
wave functions $\psi$  expressed as a  product between  quantum, quasiclassical (e.g. 
mean-field)  and classical (action phase-factor) components.  The quantum transport equation 
derived  in \cite{haat} is also  improved  by the non-linear friction term to get a  
Fokker-Planck equation,  and integrated numerically  for the two-level system. The results 
show the advantage of the  non-linear Liouville-Langevin  equation, because the presumed 
non-linear friction term in the quantum Fokker-Planck equation  does not ensure 
thermalization. 

Attempting to find a random force which could simulate quantum fluctuations,  I have 
arrived at the unexpected result  that there is no such force, but instead  that a certain 
"granularity"  is required \cite{cpw}, like in the old quantum mechanics. Thus, while the 
random force destroys the classical coherence expressed by the Hamilton-Jacobi equation, 
discretization at a certain scale may induce the "quantum" type of coherence.   

The next section reminds the framework of statistical mechanics in which the 
"classical" coherent distributions are defined, Section 3 outlines the transition to 
Wigner functions, while the Fokker-Planck equation is discussed in Section 4.    

\section{The Liouville equation}
Let $(M_\mu, \omega_\mu)$  be the phase space of a classical elementary system $\mu$  
(molecule)   with $n$ degrees of freedom \cite{somm}, and  $(M_\Gamma, \omega_\Gamma) $  
the $2 n N $ -  dimensional phase space of the ensemble $\Gamma$ (gas) consisting of 
$N$ identical elementary subsystems,
\begin{equation}
 M_\Gamma = M_1 \times M_2 \times ...\times M_N~~,~~ \omega_\Gamma =  \sum_{\mu =1}^N  \omega_\mu ~~.   \label{mgam}   
\end{equation} 
In particular, the state of a system composed of $N$ identical point-like particles 
is described  on the $6N$ dimensional manifold $M_\Gamma \equiv T^* {\mathbb R}^{3N}$ 
by a representative ("phase")  point $m$ of coordinates $(Q, P)$.
To obtain a statistical description of the ensemble each manifold $M_\mu$  is divided 
in  $K$ infinitesimal cells $\{ b_j \subset M, b_j \ne \emptyset~; b_i \cap b_j \vert_{i \ne j}  = \emptyset~,  
 i, j \in I_b \}$, $I_b = \{ 1,K \}$,  
\begin{equation}
 M_\mu = \cup_{j \in I_b} b_j~~,  \label{pmm}
\end{equation}
of volume 
\begin{equation}
\Omega_\mu^j = \int_{\bar{b}_j} \Omega_\mu 
 ~~,~~  \Omega_\mu =  \vert \omega_\mu^3 \vert /6= d^3qd^3p   ~~.    \label{evol}
 \end{equation} 
Therefore, we also obtain a partition of the manifold  $M_\Gamma$ in  $N_B=K^N$ cells $B_j$ 
of volume  $ \Omega_\Gamma^j$, $j=1,N_B$. Denoting by $w_j$ the probability to find
the representative point $m \in M_\Gamma$ at the time  $t$ in the cell $B_j$, the ratio 
$\cp_j = w_j  / \Omega_\Gamma^j$ defines the distribution function of the 
probability density  $\cp$, normalized by 
 $$ \int_{M_\Gamma}  \Omega_\Gamma  ~\cp  =1~~,~~  \Omega_\Gamma \equiv d {Q} d {P}~~. $$
It is important to remark that to address the issues of continuity and unicity of $\cp$ it
might be necessary to consider instead of a partition  (\ref{pmm}) an indexed system of open
sets $\{ U_i, i \in I \}$ covering $M_\mu$ and a system of $q$-cochains \cite{fh}, $q=0,1$, 
 associating to each set of  $q+1$ indices  $i_0, ..., i_q$ from $I$ a function $\cp_q(i_0, ...,i_q) \in 
{\mathbb R} $ on $U_{i_0} \cap U_{i_1} ... \cap U_{i_q} $.   \\ \indent 
As the Hamiltonian flow  $F_t$  on $M_\Gamma$ preserves the volume element $ \Omega_\Gamma$, 
the probability density behaves as a perfect fluid described by the continuity (Liouville)  equation    
\begin{equation}
\partial_t \cp + {\rm L}_{{\rm H} } \cp  =0   ~~,  \label{lieq}
\end{equation}
where  ${\rm L}_{{\rm H}} \cp \equiv - \{ {\rm H} ,  \cp  \} $ is the Lie derivative 
defined by the Poisson bracket and ${\rm H}$ is the total Hamiltonian, including 
interaction terms. 
\\ \indent
The dimensionality of  $\cp$ depends on the dimension of $M_\Gamma$,  $2nN$. Because in 
general $M_\Gamma$ is not a metric space it is convenient to introduce a fundamental unit 
$\hp$ for $\omega_\mu$, such that $\hp^{N n}$ is the fundamental unit for  
 $\Omega_\Gamma$. The ratio  $\gamma_j =   \Omega_\Gamma^j / \hp^{N n}$ 
is the weight of the cell $B_j$, while $ \bar{\cp}  = \hp^{N n} \cp$ is 
dimensionless, normalized by 
\begin{equation} 
\int_{M_\Gamma} \frac{  \Omega_\Gamma}{ \hp^{N n}} ~ \bar{\cp}    =1~~. \label{nc}
\end{equation}
The expectation value of a many-body observable ${\rm A} \in {\cal F}(M_\Gamma)$ (smooth 
function on $M_\Gamma$), defined by
$$ < {\rm A} > =   \int_{M_\Gamma} \frac{  \Omega_\Gamma}{ \hp^{Nn}} ~ \bar{\cp}~ 
{\rm A}~~, $$
evolves in time according to
\begin{equation} 
\frac{d <{\rm A}>}{dt} = < \{ {\rm A}, {\rm H} \} >~~.   \label{da} 
\end{equation} 
The expectation value of  $- k_B \ln \bar{\cp}$, where $k_B$ is the  Boltzmann constant, 
defines the entropy 
\begin{equation}
 \cs = -   k_B \int_{M_\Gamma}  \frac{\Omega_\Gamma}{ \hp^{N n}} ~  \bar{\cp} \ln 
\bar{\cp} ~~. \label{entr} 
\end{equation}
The one-particle probability density $\rho $ (or $\bar{\rho} = \hp^n \rho$) 
on the phase-space  $M_\mu$ is related to the density $\cp$ on $M_\Gamma$ by the
projection given by integration over $N-1$ manifolds $M_\mu$,       
\begin{equation}
\rho ( {\bf q},{\bf p} ) =  \int d^3 q_2...d^3q_N d^3 p_2 ...d^3p_N  \cp ( {\bf q}, {\bf q}_2,...,{\bf q}_N, {\bf p}, 
{\bf p}_2...{\bf p}_N ) ~~.  \label{mugd}
\end{equation}
This is well defined because the particles are identical, and although the permutations 
of coordinate indices $1,2,...N \rightarrow \{ i_1,i_2,...,i_N \}$ yield different
phase points,  $\cp$ remains invariant. For instance, if $\cp$ is a symmetric functional
$$\cp ( Q, P ) = \frac{1}{ N!  } 
\sum_{ \{ i_1 ... i_N \}} \rho_1({\bf q}_{i_1},{\bf p}_{i_1})  
 \rho_2({\bf q}_{i_2},{\bf p}_{i_2})  ...  \rho_N({\bf q}_{i_N},{\bf p}_{i_N}) $$ 
of $L \le N$ distinct distribution functions $\rho_\lambda$, $\lambda =1,L$ on $M_\mu$, then 
\begin{equation}
\rho ( {\bf q},{\bf p} ) = \frac{1}{N} \sum_{\lambda =1}^L N_\lambda \rho_\lambda ( {\bf q},{\bf p} )~~, 
\end{equation}
where $N_\lambda$ is the number of particles assigned to $\rho_\lambda$. \\ \indent
The ensemble of identical particles can also be described using the Boltzmann 
representation of "occupation numbers" in the $\mu$-space.
Thus, on $M_\mu = T^* {\mathbb R}^3$, each particle is represented by a point of coordinates 
$({\bf q},{\bf p})_i$, $i=1,N$. Let  (\ref{pmm})  be a partition of $M_\mu$  
and $N_j$ the average number of particles localized in 
the cell $b_j$.  The ratio $ {\sf f}_j = N_j /   \Omega_\mu^j$  defines the distribution function of 
the particle density ${\sf f}= N \rho$ on $M_\mu$, normalized by
\begin{equation}
\int_{M_\mu} \Omega_\mu ~{\sf f}   =N ~~. \label{nc1}
\end{equation}
If there are no interactions between  particles, ${\sf f}$ 
satisfies the one-particle Liouville equation   
\begin{equation}
\partial_t {\sf  f} + {\rm L}_H {\sf  f} =0   \label{lioeq}
\end{equation}
where $H \in {\cal F}(M_\mu) $ is the one-particle Hamiltonian and ${\rm L}_H f \equiv - 
\{ H , f \} $.  If $M_\mu =T^*{\mathbb R}^3$, then  
\begin{equation}
{\rm L}_H = (\nabla_{\bf p}  H) \cdot  \nabla - (\nabla H) \cdot  \nabla_{\bf p}   ~~,
\end{equation}
where $\nabla_{\bf p}  \equiv \vec{\partial}_p$, $\nabla \equiv \vec{\partial}_q$, 
and for 
\begin{equation}
H ({\bf q}, {\bf p}) = \frac{{\bf p}^2}{2m} + V({\bf q}) \label{h0}
\end{equation}
(\ref{lioeq}) becomes
\begin{equation}
\frac{ \partial {\sf f}}{\partial t} + \frac{{\bf p}}{m}  \cdot \nabla
{\sf f} - \nabla V \cdot \nabla_{\bf p} {\sf f} = 0 ~~.                \label{leq0}
\end{equation}
\subsection{Classical coherent states}
To solve  (\ref{leq0})  it is convenient to use the Fourier transform 
$\tilde{\sf f}({\bf q},{\bf k},t)$ in momentum,
\begin{equation}
\tilde{\sf f}({\bf q},{\bf k},t) \equiv \int d^3 p ~\re^{\ri {\bf  k} \cdot 
{\bf p}}{\sf  f}({\bf q},{\bf p},t) ~~, \label{fk1}
\end{equation}
which is a density on the configuration space ${\mathbb R}^3$ related to the particle 
(${\sf n}$) or current (${\bf j}$) densities by
\begin{equation}
{\sf n}({\bf q},t) \equiv \int d^3p ~ {\sf f}({\bf q},{\bf p}, t)  = 
\tilde{\sf f} ({\bf q},0,t)~~,  \label{n}
\end{equation}
\begin{equation}
{\bf j}({\bf q},t)  \equiv  \int d^3p ~ \frac{\bf p}{m} {\sf f}({\bf q},{\bf p}, t)  = 
- \frac{\ri}{m} \nabla_{\bf k}  \tilde{\sf f}({\bf q},0, t) ~~. \label{j}
\end{equation}
Thus, if  ${\sf  f}({\bf q},{\bf p},t)$ is a solution of (\ref{leq0}), then its Fourier 
transform  $\tilde{\sf  f}({\bf q},{\bf k},t)$ will satisfy 
\begin{equation}
\partial_t \tilde{\sf f} - \frac{\ri}{m}  \nabla_{\bf k}  \cdot \nabla  \tilde{\sf f} + \ri 
 {\bf k} \cdot (\nabla V) \tilde{\sf f} =0~~. \label{fle}
\end{equation}
An important class of solutions for the one-particle Liouville equation (\ref{leq0})  
is  represented by the   "action distributions"   
\begin{equation}
{\sf f}_0 ({\bf q},{\bf p},t) = {\sf n} ({\bf q} ,t) \delta({\bf  p}- \nabla S({\bf q},t) )
~~. \label{cs1}
\end{equation}
These are coherent functionals in the sense that remain all the time 
a product between ${\sf n} ({\bf q} ,t)$ and $ \delta({\bf  p}- \nabla S({\bf q},t) )$. 
The two real functions of coordinates and time, ${\sf n} ({\bf q} ,t)$ and $ S({\bf q},t)$ 
are related by the Hamiltonian flow because for  
\begin{equation}
\tilde{\sf f}_0 ({\bf q},{\bf k},t) = {\sf n} ({\bf q} ,t) \re^{\ri {\bf k} \cdot  
\nabla S({\bf q},t) }    \label{f0}
\end{equation}
 (\ref{fle}) reduces to the system of equations  
\begin{equation}
\partial_t {\sf  n} = - \nabla {\bf  j}        \label{co0} 
\end{equation}
\begin{equation}
{\sf n} \nabla [ \partial_t S + \frac{(\nabla S)^2}{2m} +V]=0 \label{hj}
\end{equation}
where  ${\bf j} \equiv {\sf n} \nabla S /m$ is the current density (\ref{j}). Thus, 
presuming the existence of a "momentum potential" $S({\bf q}, t)$ we get both the 
continuity and  Hamilton-Jacobi equations.
\\ \indent
In general the solutions of (\ref{hj}) are multi-valued, and ${\sf f}_0$ is a sum  
$${\sf f}_0 = \sum_i {\sf n}_i \delta({\bf  p}- \nabla S_i ) $$
over different branches. The solutions $\tilde{\sf f}_0 \equiv 
{\sf n}^{[S]}$  corresponding to the 
same function $S$ satisfy the  superposition principle,   
$({\sf n}_1+ {\sf n}_2)^{[S]} = {\sf n}_1^{[S]}+{\sf n}_2^{[S]}$,
and can be called "action waves". \\ \indent
If  ${\sf n}({\bf q})$ is a solution of the system  
(\ref{co0}), (\ref{hj}), then  $-{\sf n}({\bf q})$ is also a solution. To obtain only 
positive solutions it is convenient to search ${\sf n}= \tilde{\sf f} \vert_{{\bf k}=0}$ of the 
form ${\sf n} = \vert \psi 
\vert^2$, where $\psi$ can be a complex function. When $\psi = \sqrt{\sf n} 
\exp( \ri S / \sigma )$,  with $\sigma$ a dimensional constant,  then
\begin{equation}
\hat{\omega} \equiv   \int d^3q ~(d{\sf  n} \wedge dS) = - \ri \sigma \int d^3q ~
(d \psi^* \wedge d \psi)   
\end{equation} 
is the symplectic form induced by the complex structure of the Hilbert space 
${\cal H}= \{ \psi \in {\cal L}^2({\mathbb R}^3) \} $. The constant  
$\sigma $ and $\hp$ from (\ref{nc}) have both dimensionality of action, and in the quantum
theory $\sigma = \hp / 2 \pi \equiv \hbar$, where $\hp$ is the Planck constant. 
\\ \indent
It is important to remark that while the singularity of ${\sf f}_0$ is necessary for 
coherence, it yields infinite entropy, and therefore is not realistic. To obtain finite 
entropy we can replace for instance the delta function $\delta({\bf  p}- \nabla S)$ by a 
Gaussian\footnote{Gaussian distributions usually describe fluctuations around a mean.}  
$$g ({\bf  p}) = \frac{1}{(\pi b)^{3/2}} \re^{- ({\bf p}- \nabla S)^2/ b} 
~~,$$
where $b$ is a finite constant, but in general ${\sf f}={\sf n} g$ is not coherent. 
However, in the particular case of the harmonic oscillator  
($V({\bf q})= m \omega^2 {\bf q}^2/2$) we can find coherent solutions of the form 
$\rho_G ({\bf q},{\bf p}) = g^X ({\bf q}) g^Y ({\bf p})$,  
$$
g^X ({\bf q}) = \frac{1}{(\pi a)^{3/2}} \re^{- ({\bf q} - {\bf  X})^2/ a} ~~,~~
g^Y ({\bf p}) = \frac{1}{(\pi b)^{3/2}} \re^{- ({\bf p}- {\bf Y})^2/ b} 
~~,$$
if $b/a = m^2 \omega^2$ and ${\bf X}, {\bf Y}$ are time-dependent variables which satisfy 
the classical equations of motion, $\dot{\bf X} = {\bf Y}/m$, $\dot{\bf Y} = - m \omega^2 {\bf X}$. 
These are solutions  of constant entropy (classical, $\cs ( \rho_G) = -k_B <\ln \bar{\rho}_G> = 3 k_B 
(1- \ln 2)$, and quantum,  \cite{agar}), which can also be written in the form 
 \begin{equation}
\rho_G ({\bf q},{\bf p}) = \frac{1}{(2 \pi)^3} \int 
d^3 k ~{\re}^{- {\ri} {\bf k} \cdot {\bf p}} ~~\psi_G ({\bf q} + \frac{\sigma {\bf k}}{2})
\psi^*_G ( {\bf q} - \frac{\sigma {\bf k}}{2}) ~~,~~ \sigma = \sqrt{ab}   \label{gcs}
\end{equation}
where $\psi_G({\bf q}) = \sqrt{g^X({\bf q})} {\re}^{\ri {\bf q} \cdot {\bf  Y}/ \sigma}$. If 
$\sigma = \hbar$ then $\psi_G$ 
are (up to a phase factor) the nonstationary solutions of the TDSE known as 
Glauber coherent states. An application of such states to describe "preformed" alpha 
particles in heavy nuclei, with relevance for the Geiger-Nuttall law \cite{ab},  was 
presented in \cite{prc93}. At astronomic scale, we may presume that for a suitable 
constant $\sigma$ similar considerations might explain the "preformation" of planets 
along the orbits described by the Titius-Bode law. 
     
\section{Discretization, coherence and quantization}
The partial derivative  ${\bf k} \cdot  \nabla  S({\bf q} ,t)$ in  (\ref{f0})  is the  
limit of  
\begin{equation} 
\frac{k}{\ell} [ S({\bf q}+ \frac{\ell}{2k}  {\bf k},t)-S({\bf  q} - \frac{\ell}{2k} 
{\bf k},t)]~~,  \label{fd} 
\end{equation}
with  $k= \vert {\bf k} \vert \ne 0 $, when $\ell \rightarrow 0$. However, if 
$k \rightarrow 0$ too, a more detailed discussion might be necessary.  \\ \indent
From the early days of differential calculus, it was presumed that in (\ref{fd})
$\ell$ can be arbitrarily small, but finite. It seems though that for microparticles there 
is a physical limit  $\ell_0$, and  $\ell \rightarrow  \ell_0  >0$. The existence of an 
elementary length $\ell_0 >0$, proposed by W. Heisenberg ($\ell_0 \sim 10^{-15}$ m) and  
M. Planck ($\ell_0 \sim 10^{-34}$ m), was developed in the framework of general relativity 
theory, by the model of crystalline lattice of the physical space-time \cite{tetr}, or in  
string theory \cite{ven}. 
Independently of these considerations, the assumption of a finite limit $\ell_0$, 
expected for each massive particle near its Compton wavelength ($\ell_0 \sim 1/m$), 
was used in  \cite{cpw, rpw, gfqp} to justify the transition from a classical coherent 
distribution (\ref{f0}) to a "quantum" distribution of the form (\ref{gcs}). Let us 
presume that in the action wave (\ref{f0})  we approximate the phase ${\bf k} \cdot  \nabla  S({\bf q} ,t)$
by (\ref{fd})  and the amplitude\footnote{This means $\ln {\sf n} \approx (\ln {\sf n}_+ + \ln {\sf n}_-)/2$, 
which indicates that  such an  ${\sf n}({\bf q})$  is up to the factor $N$ a "pure" 1-particle probability 
density rather than an ensemble average.} ${\sf n}({\bf q})$  by  $\sqrt{{\sf n}_+ {\sf n}_-}$,
$$ {\sf n}_\pm={\sf n}({\bf q} \pm \frac{\ell}{2k} {\bf k})~~.$$  
Thus, the space derivative  $\partial_{q_i} S({\bf q},t) $   is replaced by the finite differences 
expression with respect to a minimum length $\ell_i = \ell k_i/k$, while the 
integration on ${\bf k}$  is limited by the size of the domain in which  ${\sf  n} ({\bf q},t) \ne 0$. 
In terms of the  new parameter $\sigma = \ell / k$, if $k \ne 0$ then      
\begin{equation}
\tilde{\sf f}_0({\bf q} ,{\bf k},t) = \lim_{\sigma \rightarrow 
0} \tilde{\sf f}_\psi({\bf q} ,{\bf k},t)  \label{lim00}
\end{equation}
where
\begin{equation}
\tilde{\sf f}_\psi ({\bf q} ,{\bf k},t) \equiv \psi^*({\bf q} 
- \frac{\sigma {\bf k}}{2},t) \psi ({\bf q} + \frac{\sigma 
{\bf k}}{2},t) \label{lim0}
\end{equation}
with  $\psi = \sqrt{{\sf n}} \exp( \ri S / \sigma )$.  Moreover, in the limit  $k \rightarrow 0$ 
$$  S({\bf q} \pm  \frac{\sigma_0}{2}  {\bf k} ,t) = S({\bf 
q},t) \pm  \frac{\sigma_0}{2}  {\bf k} \cdot \nabla 
S({\bf q} ,t) + \frac{\sigma_0^2}{8} ({\bf k} \cdot \nabla)^2  S({\bf q} ,t) \pm ...
$$
and if the terms containing  $(\sigma_0 k)^m$, $m \geq 3$, are neglected, then  
$${\bf k} \cdot \nabla S({\bf q} ,t) \approx  
\frac{1}{\sigma_0} [ S({\bf q}+  \frac{\sigma_0 }{2} 
{\bf k},t)-S({\bf q} - \frac{ \sigma_0}{2} {\bf k},t)]  $$
for any dimensional constant $\sigma_0 >0$. Therefore, within a suitable domain for  ${\bf k}$, 
we may consider $\sigma$ from (\ref{lim00}),(\ref{lim0}) as a finite constant, related 
eventually to the size of the cells $b_j$ used in the partition (\ref{pmm}). If $\sigma = 
\hbar$ then ${\sf f}_\psi$ obtained inverting (\ref{fk1}),    
\begin{equation}
{\sf f}_\psi({\bf q},{\bf p},t)= \frac{1}{(2 \pi)^3} \int 
d^3k ~{\re}^{- \ri {\bf k} \cdot {\bf p}} ~~\tilde{\sf f}_\psi ({\bf q} ,{\bf k},t) 
\label{ift}  
\end{equation}
is the Wigner transform  \cite{wig, gut} of the complex "wave function" $\psi = \sqrt{{\sf 
n}} \exp( \ri S / \sigma )$. Some properties of this functional are summarized below: \\
\noindent
- ${\sf f}_\psi$ is not positive definite, and in general it cannot represent particle 
density. However it is integrable, and the normalization 
condition  (\ref{nc}) takes the form  
\begin{equation}
\int d^3q d^3p ~~{\sf f}_\psi ({\bf q},{\bf p},t) = 
\int d^3q ~~ \vert \psi ({\bf q} ,t) \vert^2  \equiv  \langle \psi 
\vert \psi \rangle  =  N ~~,  \label{norm}
\end{equation} 
where $N$ can be seen as the number of particles (or identical 1-particle systems) used to 
define the  probability density $\rho = {\sf n} /N$.  \\
\noindent
- The "overlap"  integral  between two distributions ${\sf f}_{\psi_1}$,  
${\sf f}_{\psi_2}$, over the phase-space is   \cite{rpw}
\begin{equation}
({\sf f}_{\psi_1}, {\sf f}_{\psi_2} ) \equiv  \int  d^3q d^3p~~ 
{\sf f}_{\psi_1} {\sf f}_{\psi_2} = \frac{N}{\hp^3} < \bar{\sf f}_{\psi_2} > 
\vert_{{\rho}_{\psi_1}} = 
\frac{ \vert \langle \psi_1 \vert \psi_2 \rangle \vert^2}{(2 \pi \sigma)^3} \label{over}
\end{equation} 
where
\begin{equation}
\langle \psi_1 \vert \psi_2 \rangle \equiv \int d^3q~~ 
\psi_1^*({\bf q} ,t) \psi_2({\bf q} ,t)   \label{ampl}
\end{equation}
is the scalar product between  $ \psi_1$ and  $\psi_2$ as elements of the quantum Hilbert 
space ${\cal H}$. Thus, the overlap (\ref{over})  is positive and directly related to the  
statistical interpretation of the scalar product in quantum mechanics, suggesting again 
the choice $\sigma = \hbar$. In particular, the overlap between ${\sf f}_\psi$ and
the Gaussian (\ref{gcs}) is positive.  \\ 
\noindent
- The expectation value of a classical  observable $A$ such as 
${\bf q}$, ${\bf p}$, ${\bf p}^2$ ,  ${\bf L} = {\bf q} \times {\bf p}$,  
$$ < A >_{{\sf f}_\psi} =   \int d^3q d^3p ~{\sf f}_\psi~ A = \int d^3q~~ 
\psi^*({\bf q} ,t) \hat{A}   \psi ({\bf q} ,t) \equiv \langle  \psi \vert  \hat{A}  \vert  
\psi \rangle = \langle \hat{A} \rangle_\psi $$
is just a matrix element of the usual operator $\hat{A}$ on ${\cal H}$ associated to the 
observable $A$: $\hat{\bf q} = {\bf q}$, $\hat{\bf p} = - \ri \sigma \nabla $, 
$\hat{\bf p^2} = -  \sigma^2 \Delta $, $\hat{\bf L} = - \ri \sigma {\bf q} \times \nabla$.
For the Hamiltonian (\ref{h0})  $<H>_{{\sf f}_\psi} = \langle \hat{H} \rangle_\psi $, 
$\hat{H}=  - \sigma^2 \Delta /2m + V$, and the energy density becomes
\begin{equation} 
w_q = \int d^3 p {\sf f}_\psi H = \frac{\sigma^2}{2m} [ \vert \nabla \psi \vert^2 - 
\frac{1}{4} {\rm div} ( \psi^* \nabla  \psi + \psi \nabla  \psi^*) ] + V \psi^* \psi  ~~.
 \label{wq} 
\end{equation}
\noindent
- A symplectic diffeomorphism $\Phi$ of $(M, \omega)$ which acts on ${\sf f}_\psi \in {\cal F}(M)$ 
by $\Phi^* {\sf f}_\psi = {\sf f}_{\psi'}$ yields, according to (\ref{over}), a unitary transformation 
$ \hat{U}_\Phi$  of the state vectors $\psi \in {\cal H}$ of the form  $\psi ' = 
\hat{U}_\Phi^{-1} \psi$, such that 
\begin{equation}
\Phi^* {\sf f}_\psi ={\sf  f}_{\hat{U}_\Phi^{-1} \psi}  ~~.  \label{dsy}
\end{equation} 
In particular, when $\Phi$ is the action of a Lie group $G$, the infinitesimal 
transformations take the form $\hat{U}_\epsilon = 1 + \ri \epsilon \hat{J}$, where $\hat{J}$ 
are Hermitian operators associated to the elements of the Lie algebra ${\mathfrak g}$ of $G$. 
Thus, the main difference between the classical and quantum outcome of dynamical symmetries 
(e.g. at spontaneous symmetry breaking) is due to their realization within spaces having 
different coherence  properties (${\sf f}_0$ and ${\sf f}_\psi$).  \\ \indent
 In general, a  functional   ${\sf f}_{[{\sf n},S]}  $ of ${\sf n}$ and $S$, will be called 
coherent with respect to the classical Liouville equation if during time-evolution it 
remains the same functional, although ${\sf n}$ and $S$ may change. According to \cite{cpw},
if the potential in $H$  is a constant, linear, or quadratic polynomial of  
$q$, then ${\sf f}_\psi$ is an exact solution of the Liouville equation  
$\partial_t {\sf f}_\psi =  \{ H, {\sf f}_\psi \}$  if $\psi$  is an exact solution of 
\begin{equation} 
\ri \sigma \partial_t \psi = \hat{H} \psi~~,~~
\hat{H}=  - \frac{\sigma^2}{2m} \Delta + V ~~, \label{se}
\end{equation}
formally identical to the time dependent Schr\"odinger equation (TDSE). Thus, ${\sf f}_0$ is 
coherent for any Hamiltonian, but  ${\sf f}_\psi$  is coherent only for polynomial 
potentials of degree at most 2.  This restriction  was derived  before 
using different arguments  both in algebraic and geometric quantization \cite{am, gfqp}.  
\\ \indent
Beside the stability of the functional form, another aspect of the "coherence" property 
is  that for  ${\sf f}_0$ and  ${\sf f}_\psi$ the two functions ${\sf n}$ and $S$ 
play also the role of canonically conjugate variables. This aspect is particularly 
important for real waves quanta such as photons or phonons \cite{ldb}, and it can be shown 
\cite{cpw}  that the equations of motion for these variables can be derived from a 
variational  principle related to infinite-dimensional Hamiltonian systems of the form   
\begin{equation}
 i_{X_W} \hat{ \omega} =  d W ~~, \label{idhs}
\end{equation}
 where 
$ \hat{ \omega} = d \hat{\theta}$, 
$$
\hat{ \theta} = \int d^3 q~  {\sf n}  d S~~,~~X_{W} = \int d^3 q~ 
(\partial_t {\sf n} \frac{\partial}{\partial {\sf n}} + \partial_t {S} 
\frac{\partial}{\partial S } )~~,
$$
and  $W = \int d^3 q~  w $ is the classical ($w \equiv w_{cl} = 
{\sf n} H(\nabla S, {\bf q})$) or quantum, ($w \equiv w_q $, (\ref{wq}) ) energy functional of 
${\sf n}$ and $S$. \\ \indent
For an integrable  distribution ${\sf f} \in L^1(M)$ on the symplectic manifold 
$M=T^*Q$, the  "coordinates" $({\sf n},  S )$  presume a foliation of $M$ by Lagrangian  
submanifolds  $\Lambda_S \subset M$  generated by $S \in C^1(W), W \subset Q $, and  the 
projection 
$$ \pi : L^1(M) \mapsto L^1 (Q)~~,~~\pi ({\sf f}) = {\sf n}$$
defined  by  integration on $\Lambda_S$.  If $S$ is the solution of the Hamilton-Jacobi 
equation, then the  asymptotic solution of the Schr\"odinger equation  (\ref{se}) in the 
WKB approximation $\psi  \sim  \exp ( {\rm i}  S / \sigma)$ is related to the subspace of 
polarized sections $r \in \Gamma_L(M,\Lambda_S)$ autoparallel on $\Lambda_S$  
( $\nabla_X r=0, ~\forall X \in T \Lambda_S$)  in a complex Hermitian line-bundle with 
connection  $(L, \nabla)$ over $M$ \cite{gfqp, nw, bk}. Thus, the  subspace of the coherent 
functionals  ${\sf f}_\psi$ defined by  the Wigner transform arises by a peculiar lift of
$\Gamma_L(M,\Lambda_S)$ to $\Gamma_L(M)$.  \\  \indent
Although the relationship between  $\psi$ and ${\sf f}_\psi$ is nonlinear, and 
${\sf f}_{\psi_1 + \psi_2}  \ne  {\sf f}_{\psi_1} + {\sf f}_{\psi_2}$,
we note that if   $\psi_1$, $\psi_2$  are solutions of TDSE, and  
${\sf f}_{\psi_1}$,  ${\sf f}_{\psi_2}$ satisfy the Liouville equation, then 
${\sf f}_{\psi_1 + \psi_2 }$ is also a solution of the Liouville equation\footnote{The spectacular 
case in which the interference terms between $\psi_1$  and $\psi_2$ may produce the resonant 
transfer of a heavy particle across a macroscopic distance is discussed in  \cite{ldqt} using estimates 
based on a finite triple-well potential.}.  Denoting by $\hat{P}_\psi \equiv \vert \psi \rangle 
\langle \psi \vert$  the projection operator  associated to the state function $\psi$, 
$\langle \psi \vert \psi \rangle =1$, the distribution ${\sf f}_\psi$ (\ref{ift}) takes 
the form
\begin{equation}
{\sf f}_\psi ({\bf q}, {\bf p} ) =  {\sf W} (\hat{P}_\psi ) \equiv  \frac{1 }{(2 \pi)^3} 
\int  d^3 k  {\rm e}^{- {\rm i} {\bf k} \cdot {\bf p}} \langle {\bf q } \vert  
\hat{U}_{k/2}  \hat{P}_\psi  \hat{U}_{k/2} \vert {\bf q} \rangle  \label{wop} 
\end{equation}
with $\hat{U}_{k/2} = \re^{\ri {\bf k} \cdot \hat{\bf p} /2}$. Thus, between  
${\sf f}_\psi$ and $\hat{P}_\psi$ there exists a linear relationship by the transform  
${\sf W}$. Moreover, if ${\cal C}$ denotes a complete set of states, ($\sum_{\psi \in {\cal C}}  \hat{P}_\psi = \hat{1}$),  then 
\begin{equation}
\sum_{\psi \in {\cal C}} {\sf f}_\psi ({\bf q}, {\bf p} ) = \frac{1 }{(2 \pi \sigma )^3 } ~~. 
\label{sum} 
\end{equation}
In terms of group actions, the configuration space $Q={\mathbb R}^3$ is homogeneous space 
for the Lie group $G={\mathbb R}^3$ of the space translations, $T_q Q  \simeq T_eG \equiv 
{\mathfrak g}$, the momentum space $T^*_qQ$ is parameterized by  ${\bf p} \in 
{\mathfrak g}^*$, and ${\bf k}$, the Fourier dual to ${\bf p}$, enters in  (\ref{wop}) as 
a parameter on $G$.   
\\ \indent  
The eigenfunctions $\psi_\lambda$ of $\hat{H}$,    
\begin{equation}
\hat{H} \psi_\lambda = E_\lambda \psi_\lambda          \label{esst}
\end{equation}   
are stationary solutions of  (\ref{se}),  $\psi_\lambda (t) = \re^{- i E_\lambda  t/ \sigma } 
\psi_\lambda (0)$, and correspond to distributions ${\sf f}_{\psi_\lambda} $  independent of time,
of  energy   
\begin{equation}
E = < H > \vert_{{\sf f}_{\psi \lambda}} =   \int d^3q d^3p ~{\sf f}_{\psi_\lambda}  H =  
\langle  \psi_\lambda \vert  \hat{H}  \vert  \psi_\lambda \rangle = E_\lambda~~.  \label{eav} 
\end{equation} 
It is important to remark that this equality, which is used in many stationary 
variational calculations, holds for any Hamiltonian of the form (\ref{h0}) \cite{sd}.  

\subsection{ Relativistic Wigner functions and Schr\"odinger equation} 
The problem of relativistic Wigner functions and Schr\"odinger equation for massive 
particles was studied  in \cite{rpw} within  the extended phase-space 
$M^e=T^*{\mathbb R}^4$ presented in \cite{et}. Thus,  the energy ($E$) and time 
($t \equiv q_0/c$) become conjugate variables, evolving with respect to a true parameter 
$u$, called universal time. 
A particular class of coherent solutions for the relativistic Liouville equation 
(RLE) consists of the "action distributions"  
\begin{equation}
f_0^e(q^e,p^e,u) = n^e(q^e,u) \delta( p_0 - \partial_0 S ) \delta( {\bf p} - 
\nabla S ) ~~, 
\end{equation}
where $n^e$ is the probability density of  localization in space-time. Considering 
$\partial_u S = m_0c^2$, in the case of a free particle we get the continuity equation
\begin{equation}
m_0 \partial_u n^e = \partial_0 (n^e \partial_0 S) -  \nabla \cdot (n^e \nabla S)~~, 
\label{cont1}
\end{equation}
and 
\begin{equation}
(\partial_0 S)^2 - (\nabla S)^2 = m_0^2 c^2~~.   \label{rhj}
\end{equation}
For a density   
$n^e(q^e,u) = \delta (q_0-c u) n({\bf q},u)$, localized in time, (\ref{cont1}) reduces 
in the nonrelativistic limit to the usual continuity equation 
$\delta (t -u) [m_0 \partial_u n +  \nabla \cdot (n \nabla S)] =0$. The 
nonrelativistic identification of  $u$ as time may appear when $t$  is a quasiclassical 
variable \cite{vpr}, described by a Gaussian wave-packet such that $\langle t \rangle 
=u$.  The width of this wave-packet  sets a lower limit for the classical time-intervals, 
and a fundamental space-length $\ell_0$. Evidence for the existence  of such an elementary 
time-interval  $\delta t_0 = \ell_0 /c= \hbar /m_0c^2$ was found in the particle 
data  \cite{rpw}.
\begin{figure}
\begin{picture}(250,100)(0,100)
\put(-20,-70){\includegraphics[width=5in,height=5in]{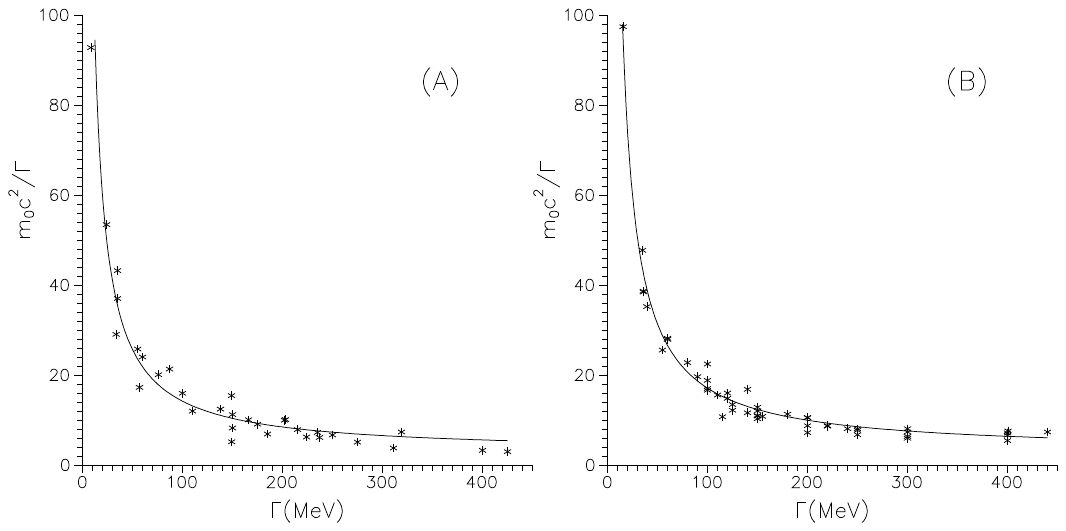}}
\end{picture} \\  {\small  Figure 1. $ m_0c^2 / \Gamma$ from experimental data ($*$) and the 
interpolation functions $2.1 + C /\Gamma$ (solid) for 32 light unflavored meson resonances 
($\omega, \eta, \phi, \pi, \rho, a,b,f$) 
with $\Gamma \ge 8.43$ MeV (A) and 48 baryon resonances ($N,\Delta ,\Lambda , \Sigma$) 
with $\Gamma \ge 15.6$ MeV (B) \cite{rpw}.   }    
\end{figure}
Thus, the ratio $ m_0c^2 / \Gamma =  \tau_L / \delta t_0$ between the mass (in MeV) and 
decay width ($\Gamma$), calculated using the experimental data\footnote{Some of 
these old data do not appear anymore in \cite{data}.} for meson and baryon resonances 
is well interpolated by functions of the form $2.1 + C /\Gamma$, where 
$C$ is 1222 MeV for mesons and  1487 MeV for baryons (Figure 1, Appendix 1), indicating that 
the lifetime $\tau_L = \hbar / \Gamma$ is  limited below by $2  \delta t_0 $.   \\ \indent
A quantum distribution 
\begin{equation}
\tilde{f}^e_\Psi (q^e,k^e,u)  \equiv \Psi (q^\mu + \frac{\sigma k^\mu}{2},u) \Psi^*(q^\mu 
- \frac{\sigma k^\mu}{2},u) ~~,   \label{fq}
\end{equation} 
is a "static" solution ($\partial_u \tilde{f}^e_\Psi =0$) of the RLE  if 
$- \sigma^2 \Box \Psi = m_0^2c^2 \Psi$, $\Box \equiv \partial_0^2 - \nabla^2$. When 
$\sigma = \hbar$ this represents the Klein-Gordon equation. \\ \indent
The extended phase-space is also the suitable framework to describe the electromagnetic 
field  \cite{em2f}. In vacuum the electric and magnetic fields ${\bf E}$ and ${\bf B}$ 
appear as coefficients of two dual 2-forms $\omega_f, \omega^*_f$ on the space-time 
manifold ${\mathbb R}^4$,  
\begin{equation}
\omega_f =  - {\bf B} \cdot d {\bf S} + {\bf E} \cdot d q_0 \wedge  d {\bf q}~~,~~
\omega^*_f = {\bf E} \cdot d {\bf S} +  {\bf B} \cdot d q_0 \wedge d {\bf q}~~,
\end{equation} 
where $dS_1 = dq_2 \wedge dq_3$, $dS_2 = - dq_1 \wedge dq_3$, $dS_3 = dq_1 \wedge dq_2$.
In the presence of the field the canonical symplectic form $\omega_0^e$ on 
$T^*{\mathbb R}^4$ for a relativistic massive particle which carries the 
electric charge $q_e$ and the magnetic charge $q_m$ becomes \cite{em2f} 
\begin{equation}
\omega^e = \omega_0^e + \frac{q_e}{c} \omega_f + \frac{q_m}{c} \omega_f^*~~, \label{ome}
\end{equation}      
to account for the Lorentz forces ${\bf F}_B = q_e{\bf v } \times {\bf B}/c$ and
${\bf F}_E=  - q_m {\bf v } \times {\bf E} /c$. By specific integrality conditions these 
two forms provide electric or magnetic charge quantization, while the exterior derivatives   
in vacuum $d \omega_f = d \omega_f^* =0$ yield the wave equation $\Box {\bf E} = 
\Box{\bf B} =0$. Such an equation has as coherent solutions any vector function of
$\tau = t  \pm  {\bf n} \cdot {\bf q}/ c$,  where ${\bf n}$ is the unit vector along the 
propagation direction. For instance, ${\bf E}$ can be a harmonic or a Gaussian function of
$\tau$, as we may have a plane wave or a localized pulse. However, $\tau$ is not 
Lorentz-invariant, and instead  it is convenient  to consider coherent  functionals  
of a Lorentz-invariant  phase function  $\varphi (q_0,{\bf q}) \sim \tau $. \\ \indent
The photon, as a relativistic particle of vanishing rest mass and energy $\epsilon = c \vert
{\bf p} \vert$, associated with the (real) electromagnetic waves, can be introduced 
considering the energy-density continuity equation 
\begin{equation}
\partial_t w_f = - {\rm div} {\bf Y} ~~, \label{wy}
\end{equation}
and the eikonal equation
\begin{equation}
({\bf \nabla} \varphi)^2 = (\partial_0 \varphi)^2~~, \label{eik}
\end{equation}
which are similar to (\ref{cont1}) and (\ref{rhj}) with $m_0=0$. 
Here $w_f = ( {\bf E}^2 + {\bf B}^2 )/2$ is the energy density of 
the field and ${\bf Y} = c {\bf E} \times {\bf B} $ is the Poynting vector. Because  
photons are free particles there  are  no "zero-point energy"   
terms in  the Planck distribution, (accurately retrieved in the 2.7 K cosmic  microwave 
background  spectrum \cite{b3}), or in the  vacuum energy density \cite{b4}.  \\ \indent
The case of particles in states of negative energy ($E<0$)  is peculiar  because according 
to \cite{et, rpw}, in such states the Lorentz group $SO(1,3)$ is replaced by $SO(4)$, 
locally isomorphic to $SU(2) \times SU(2)$. This means that  for $E<0$  the distinction between 
space and time coordinates disappears, and apparently such particles are 4-dimensional objects 
living in closed space-time domains. It is interesting to remark that in general 
relativity the metric outside a spherical shell of mass $M$ \cite{dirac},
$$ d s^2 = (1- 2 \gamma_G \frac{M}{r c^2} ) d q_0^2 - (1- 2 \gamma_G \frac{M}{r c^2} 
)^{-1}  d r^2  - r^2 ( d \theta^2 + \sin^2 \theta d \varphi^2)~~,   $$  
$\gamma_G = 6.67 \cdot 10^{-11}$ Nm$^2$/kg$^2$,
shows no clear distinction between space and time coordinates at $r<R_g= 2 \gamma_G M /c^2$, 
when formally the gravitational (binding)  self-energy approaches $- Mc^2$.  
Beside this similarity, we may speculate that the effect of the inertial parameter (mass) 
on the metric described at macroscopic scale by the general relativity may turn 
at atomic scale into an effect on the constant $\ell_0$.  

\section{Finite temperature effects} 
At finite temperature ($T$) the thermal noise affects the statistical ensemble of the "pure" 
action  distributions (\ref{cs1}), which evolve towards the classical equilibrium density ${\sf f}_e$, 
$${\sf f}_e  = \frac{N}{\hp^3} \frac{\re^{- \beta H  }}{Z_\mu}~~,~~
Z_\mu =   \int_{M_\mu}   \frac{\Omega_\mu}{\hp^3}  ~  \re^{- \beta H}~~,$$
$\beta = 1 / k_B T$,  according to the  Fokker-Planck equation  
\begin{equation}
\partial_t {\sf f}  + \frac{1}{m} {\bf p} \cdot \nabla {\sf f} 
 - \nabla V \cdot \nabla_{\bf p} {\sf f}  = \gamma \nabla_{\bf p} \cdot (  \frac{\bf p}{m} +
\frac{\nabla_{\bf p}}{\beta }  ) {\sf f}  ~~,   \label{fpl}
\end{equation}
where $\gamma$ denotes the friction coefficient. By the Fourier transform in momentum 
(\ref{fpl}) becomes   
\begin{equation}
\partial_t \tilde{\sf f} - \frac{\ri}{m}  \nabla_{\bf k}  \cdot \nabla  \tilde{\sf f} +  
 {\bf k} \cdot (\ri  \nabla V +  \frac{\gamma}{m} \nabla_{\bf k}  ) \tilde{\sf f} =
- \gamma k_B T k^2 \tilde{\sf f} ~~. \label{fpl1}
\end{equation}
A function $\tilde{\sf f}$ of the quantum  form $\tilde{\sf f}_\psi( {\bf q},{\bf k}) = 
\psi({\bf a}) \psi({\bf b})^*$, (Appendix 2),  with ${\bf a}= {\bf q} + \hbar {\bf k}/2$ and 
${\bf b}= {\bf q} - \hbar {\bf k}/2$, can be written as a matrix element 
$\tilde{\sf f}_{ab} = \langle {\bf a} \vert \psi \rangle \langle \psi \vert {\bf b} \rangle$ 
of the operator $\vert \psi \rangle \langle \psi \vert$ between the eigenstates
$\vert {\bf a}\rangle , \vert {\bf b} \rangle$ of the position operator $\hat{\bf q}$. 
With the notation $\hat{\sf f}_{ab} = \tilde{\sf f}_{ab} / \hp^3$ we also get $\partial_t \hat{\sf f}_{ab}  = (\partial_t \hat{\sf f})_{ab}$, 
$${\bf k} = \frac{1}{\sigma} ({\bf a} - {\bf b})  ~~,~~ {\bf k}\hat{\sf f}_{ab}=  
\frac{1}{\sigma} [\hat{\bf q}, \hat{\sf f}]_{ab}~~,~~
k^2 \hat{\sf f}_{ab} = \frac{1}{\sigma^2} [\hat{\bf q} \cdot, [{\bf q}, \hat{\sf f}]]_{ab}~~,$$ 
$$\nabla_{\bf k}  = \frac{\sigma}{2} (\nabla_{\bf a} - \nabla_{\bf b})~~,~~ 
\nabla_{\bf k}  \hat{\sf f}_{ab} =  \frac{\sigma}{2} \{ \nabla , \hat{\sf f} \}'_{ab}~~,~~
\nabla_{\bf k}  \cdot \nabla \hat{\sf f}_{ab} = \frac{\sigma}{2} [\Delta, \hat{\sf f}]_{ab}~~, $$   
where $\{  ,  \}'$ denotes the anticommutator. At small $k$  
$$\sigma {\bf k} \cdot \nabla V  \hat{\sf f}_{ab} \approx  (V_a -V_b) \hat{\sf f}_{ab}=[V, \hat{\sf f}]_{ab}~~,$$
and as indicated in \cite{cpw},  if $\sigma = \hbar$,  $\hat{\sf f}/N = \hat{\rho}= \hat{\bar{\rho} }/ \hp^3$, 
$Tr \hat{\bar{\rho}} =1$,  then for a single microscopic particle  (\ref{fpl1})  takes the form of the quantum 
Fokker-Planck equation for the usual density operator $\hat{\bar{\rho}}$,
\begin{equation}
\ri \hbar \partial_t \hat{\bar{\rho}} = [\hat{H}, \hat{\bar{\rho}}] +
 \frac{\gamma}{2 m} [{\bf q}, \cdot  \{ \hat{\bf p} ,\hat{\bar{\rho}} \}']
- \frac{ \ri \gamma k_B T}{\hbar} [\hat{\bf q}, \cdot [\hat{\bf q}, \hat{\bar{\rho}} ]] ~~.
\label{fpe}
\end{equation} 
This equation is similar to the one proposed in  \cite{cl}, and it 
takes the form considered in \cite{vpr} if $\hat{\bf p }$  from $\{ \hat{\bf p} ,\hat{\bar{\rho}} \}'$
is replaced by  $\langle \hat{\bf p } \rangle $. Though, none of them has a satisfactory form, 
apparently due to the dissipative term. In classical mechanics, the effect of dissipation 
is not only energy loss, but also  decrease in the phase-space volume. Thus, density fluctuations 
near the  volume of the elementary cell may  change the dissipation mechanism. In 
fact, when $\hat{H} = \hat{\bf p}^2 / 2m$ the quantum equilibrium distributions 
\begin{equation}
\hat{\bar{\sf f}}_\pm = \frac{1}{\re^{ \beta \hat{H} - \alpha } \pm 1} 
~~,~~Tr \hat{\bar{\sf f}}_\pm =N~~, \label{qed}
\end{equation}
can be obtained as stationary solutions of the nonlinear equation 
\begin{equation}
\ri \hbar \partial_t \hat{\bar{\sf f}} = [\hat{H}, \hat{\bar{\sf f}}] + \frac{\gamma}{2 m} [{\bf q}, 
\cdot  \{ \hat{\bf p} ,\hat{\bar{\sf f}} ( 1 \mp  \hat{\bar{\sf f}}) \}']
- \frac{ \ri \gamma k_B T}{\hbar} [\hat{\bf q}, \cdot [\hat{\bf q}, \hat{\bar{\sf f}} ]] ~~,
\end{equation} 
in which $\mp \gamma  \{ \hat{\bf p} ,\hat{\bar{\sf f}}^2 \}'/2m$ could be  assigned to a 
density-dependent correction term to the friction force  \cite{sd}. However, before thermalization, 
while the thermal noise decreases the coherence domain, one can expect a transition from 
complex ($\psi$) "probability waves" to real (${\sf n}$) density waves \cite{cpw}. 

\section{Conclusions} 
The Fourier transform in momentum $\tilde{\sf f}$ of the distribution function on the 
classical phase space is a density on the configuration space $Q$, such that coherent 
solutions of the Liouville equation, expressed as functionals of only two functions on $Q$, 
${\sf n}$ and $S$, can be found. The action waves ${\sf f}_0$ (\ref{cs1}) are localized in 
momentum and evolve according to the Hamilton-Jacobi equation at infinite entropy. 
Coherent  distributions (\ref{gcs}) of finite entropy can be found in the 
particular case of the harmonic oscillator. These are coherent not only as Gaussian 
distributions on the phase-space, but also as the Wigner transform of the Glauber states,
coherent for TDSE. In general, a functional ${\sf f}_0$ takes the form of the Wigner 
function ${\sf f}_\psi$ (\ref{ift}) by space discretization. Although ${\sf f}_\psi$ is not 
positive definite, it has a positive overlap with (\ref{gcs}), and in this sense can be 
considered as particle (or probability) density. During time evolution ${\sf f}_\psi$, with 
$\psi$ an exact solution of TDSE, remains coherent only for polynomial potentials of degree 
at most 2. This means for instance that in a Coulomb potential either the classical 
Liouville equation, or TDSE should contain correction terms, which can be calculated and 
compared to other corrections (e.g. relativistic,  QED  \cite{b2}), or experimental 
data (e.g. the transition time in single atoms \cite{qj}). However, the equality (\ref{eav})
$<H>\vert_{{\sf f}_\psi} = \langle \hat{H} \rangle_\psi $, which can be used in 
time-independent variational calculations for the dominant part of the (quasi) stationary
equilibrium distributions, holds for any potential \cite{sd}. \\ \indent
Relativistic Wigner functions can be defined similarly, by space-time discretization of the  
action distributions on the extended phase space. Though, the presumed dependence 
of the minimum interval of time (the "present") on the inertial parameter, or the 
problem of negative mass, indicate that the suitable framework for discussion is the 
general relativity.    \\ \indent
At finite temperature the Fourier transform (\ref{fpl1}) of the classical Fokker-Planck 
equation takes the form of the quantum transport equation (\ref{fpe}) simply by considering 
the density $\tilde{\sf f}_\psi( {\bf q},{\bf k})$ as a matrix element.  However, to obtain
the quantum equilibrium distributions (\ref{qed}) as stationary solutions an additional, 
density-dependent correction to the dissipative term, is necessary. \\ \indent
The results summarized above indicate that the functional coherent distributions on the 
classical phase-space may provide the missing link between classical 
mechanics and quantum phenomenology. The "action waves" (\ref{cs1})  and the Wigner 
functions (\ref{ift})  are two examples of coherent distributions ${\sf f}_{[{\sf n},S]} $  
related  to the classical and quantum behavior, respectively, but the space 
of such solutions, its relationship to "granularity", and the various aspects of 
"decoherence" remain so far unexplored. 

\newpage
\section{\bf Appendix 1:  Particle data tables }
 
{\bf Table 1.} Comparison between the experimental value of the mass ($M$) and the 
estimate $2.1 \Gamma+$ 1222 MeV for some meson resonances ($M, \Gamma$ from \cite{data}).   \\

\begin{tabular}{|c|c|c|c|c|c|c|c|}
\hline
Resonance  &  $\Gamma$ (MeV)   & $J^{PC}$   & $M$ (MeV)  & $2.1 \Gamma+$1222 MeV  \\ \hline
 $f_1$ (1285) & 24.2 $\pm 1.1$   & 1$^{++}$   & 1282.1 $\pm 0.6$   & 1272.8 $\pm 2.3$   \\ \hline
 $\eta$ (1295) & 55 $\pm 5$  & 0$^{-+}$   & 1294 $\pm 4$   & 1337.5 $\pm 10.5$  \\ \hline
 $f_0$ (1500) &   109 $\pm 7$   & 0$^{++}$   & 1505 $\pm 6$  & 1450.9 $\pm 14.7$  \\ \hline
 $\pi$ (1800) &  208 $\pm 12$   & 0$^{-+}$   &  1720 $\pm 6$  & 1658.8 $\pm 25.2$ \\ \hline
\end{tabular}
\vskip0.5cm
\noindent
{\bf Table 2.} Comparison between the experimental value of the mass ($M$) and the estimate 
$2.1 \Gamma +$ 1487 MeV  for some baryon resonances ($M, \Gamma$ from \cite{data}).  \\ 

\begin{tabular}{|c|c|c|c|c|c|c|c|}
\hline
Resonance  &  $\Gamma$ (MeV)   & $J^{P}$   & $M$ (MeV)  & $2.1 \Gamma+$1487 MeV  \\ \hline
$\Lambda$(1520) & 15.6 $\pm 1$ & $\frac{3}{2}^-$ & 1519.5 $\pm 1$ & 1519.7 $\pm 2.1$   \\ \hline
N(1700) & 150 (100-250) & $\frac{3}{2}^-$ & 1700(1650-1750)  & 1802(1697-2012) \\ \hline
$\Sigma$(1940) & 220 (150-300) & $\frac{3}{2}^-$ & 1940 (1900-1950) & 1949 (1802-2117) \\ \hline
N(2600) & 650 (500-800) & $\frac{11}{2}^-$  & 2600(2550-2750) & 2852 (2537-3167) \\ \hline
\end{tabular}
\vskip0.5cm 

\section{\bf Appendix 2: Partitions and characteristic functions}

Let (\ref{pmm}) be a partition of the $\mu$-space $M=T^*\mathbb{R}^3$ and $\chi_i : M \rightarrow \mathbb{R}$ 
the characteristic function of the cell $b_i$. If all cells have the same volume $\Omega_\mu^k = \Omega_0$, 
(the granularity), then the elements of the set
$C_b = \{ \chi_i : M \rightarrow \mathbb{R}~/~ \chi_i \vert_{b_k} = \delta_{ik},~
i,k \in I_b \} $ have the properties: \\
\begin{equation}
 1.  ~ \int_M \Omega_\mu \chi_i =\Omega_0 ~,~  
2. ~(\chi_i, \chi_j) = \int_M \Omega_\mu \chi_i \chi_j=  \Omega_0 \delta_{ij}~,~
3.~ \sum_{i \in I_b} \chi_i (m) =1~, \forall m \in M~.   \label{cbi} 
\end{equation}   
The partition is presumed adapted to a real polarization of $(M, \omega)$ such 
that the cells can be separated by Lagrangian submanifolds $\Lambda_Q$, 
$\Lambda_P$, and $\chi_i ({\bf q},{\bf p})$ is separable in the coordinates $({\bf q},
{\bf p})$ as $\chi_i({\bf q},{\bf p}) = \chi_i^Q({\bf q}) \chi_i^P({\bf p})$, $\forall i 
\in I_b$. The cells are  considered identical: cubic, simply connected, although in 
principle we may consider also quasi-degenerate partitions with elongated, 
string-like cells, or  with flat cells, almost plane. \\ \indent
An observable $f \in {\cal F}(M)$ is called macroscopic with respect to $C_b$ 
if it can be accurately approximated by $f_b = \sum_{k \in I_b} f_k \chi_k$
with $f_k = (f, \chi_k) / \Omega_0$. If $\rho$ is a macroscopic probability
distribution and $A$ a macroscopic observable then $\Omega_0 \rho_k$ is the 
probability of localization in the cell $b_k$ and 
\begin{equation}
<A>_\rho = (A, \rho) = \sum_{k \in I_b} (A, \chi_k) \rho_k = \Omega_0 
\sum_{k \in I_b} A_k \rho_k
\end{equation}
is the expectation value of $A$.  \\ \indent
If we change $C_b$ to $C_{b'} = \{ \chi_k', k \in I_{b'} \}$ by global 
translations, rotations or variations in the shape of the cells at constant 
volume then formally $\chi_i' = \sum_{k \in I_b} P_{ik}^b \chi_k$ with $P_{ik}^b =
( \chi_i', \chi_k) / \Omega_0 \ge 0$, but the matrix $P^b$ may be singular because
$\chi_i'$ are not macroscopic observables. The set $\{ P_{ik}^b,~ k \in I_b \}$ does not 
specify $b_i'$  completely, but it can be interpreted as its probability distribution over the
partition  $\{ b_k, k \in I_b \}$ , related to $\{ b_i'  \cap b_k,  k \in I_b \}$.     \\ \indent
 Let $\tilde{f}_1, \tilde{f}_2$  be the Fourier transforms in momentum of 
$f_1, f_2 \in {\cal F}(M)$. Then 
\begin{equation}
(f_1, f_2) = \int_M d^3q d^3p~ f_1({\bf q},{\bf p}) f_2 ({\bf q},{\bf p})  = 
\frac{1}{(2 \pi)^3} \int d^3q d^3k~ \tilde{f}_1({\bf q},{\bf k}) \tilde{f}_2 ({\bf q},{- \bf k}) ~~. \label{ops}
\end{equation}
Introducing new variables $({\bf a},{\bf b})$: ${\bf a} = {\bf q} + 
\hbar {\bf k}/2$,  ${\bf b} = {\bf q} - \hbar {\bf k}/2$, and the notation 
$\hat{f}_{ab}= \tilde{f} ({\bf q},{\bf k}) / \hp^3$, $\hat{f}_{ab}^* = \hat{f}_{ba}$, 
(\ref{ops}) becomes   
\begin{equation}
(f_1, f_2) = \hp^3 \int d^3a d^3b~ \hat{f}_{1ab}  \hat{f}_{2 ba}  \equiv \hp^3 Tr( \hat{f}_1 \hat{f}_2) ~~.
 \end{equation}
In particular, when $A \in {\cal F}(M)$ is $1$, $q_i$ or $p_i$ we get 
$\hat{1}_{ab} = \delta ( {\bf a} - {\bf b})$, 
$(\hat{q}_i)_{ab} = a_i \delta ( {\bf a} - {\bf b})$,
$(\hat{p}_i)_{ab} = - (\ri \hbar /2) (\partial_{a_i} - \partial_{b_i}) 
\delta ( {\bf a} - {\bf b})$. \\  \indent
A function $\chi_\psi \in {\cal F}(M)$ is called associated to a "quantum cell"
$q_\psi$ (topological subspace of $\mathbb{R}^6$) if $(\hat{\chi}_\psi)_{ab}$
is separable in the variables ${\bf a}$ and ${\bf b}$ such that  $(\hat{\chi}_\psi)_{ab} = 
\psi({\bf a}) \psi({\bf b})^*$  (or $\hat{\chi}_\psi = \vert \psi \rangle 
\langle  \psi \vert \equiv \hat{\bar{\rho}}_\psi$) with $\psi \in  
{\cal L}^2 ( \mathbb{R}^3)$, $\langle \psi \vert \psi  \rangle =1$. \\ \indent
Let ${\cal E}= \{ \psi_n \in {\cal H}  /  \langle \psi_n \vert  \psi_{n'}  \rangle = \delta_{n n'}~, 
n, n' \in I_q \}$ be a countable orthonormal basis in ${\cal H}= {\cal L}^2 
(\mathbb{R}^3)$.  Then the elements of the set $C_q = \{ \chi_{\psi_n} 
\in {\cal F}(M)~,~ \psi_n \in {\cal E} \} $ have the properties: \\    
\begin{equation}
 1.  ~ ( \chi_{\psi_n}, 1) =\hp^3 ~~,~~  
2. ~( \chi_{\psi_n}, \chi_{\psi_{n'}}) =   \hp^3 \delta_{nn'}~~,~~
3. ~\sum_{n \in I_q}  \chi_{\psi_n}(m) =1~,~\forall m \in M~,
\end{equation}  
similar to (\ref{cbi}) with $\Omega_0 = \hp^3$, while $\rho_\psi = \chi_\psi / \hp^3$ is
the Wigner transform of $\psi$. A reference set is 
$C_{\hat{H}} = \{ \chi_{\psi_n} \in {\cal F}(M)~/~\hat{H} \psi_n = 
E_n \psi_n~,~ \langle \psi_n \vert  \psi_{n'}  \rangle = \delta_{n n'},~ 
n,n' \in I_q \}$, defined by the eigenfunctions (\ref{esst}) of the
1-particle Hamiltonian $\hat{H}$. With respect to this set, a macroscopic 
(thermal) probability distribution on $M$ has the form $\rho_q = \sum_{n 
\in I_q} w_n \rho_{\psi_n}$, with $w_n \ge 0$, $\sum_{n 
\in I_q} w_n =1$, $\hp^3 \rho_{\psi_n} = \chi_{\psi_n} \in
C_{\hat{H}}$, or $\hat{\rho}_q = \hat{\bar{\rho}}_q / \hp^3$ where 
$\hat{\bar{\rho}}_q = \sum_{n  \in I_q} w_n 
\vert \psi_n \rangle \langle \psi_n \vert$ is the density operator. 
\\ \indent
For an arbitrary function  $\psi \in {\cal H}$,  $\langle \psi  \vert  \psi  \rangle =1$, the 
coefficient $P^q_{\psi \psi_n} = (\chi_\psi, \chi_{\psi_n}) / \hp^3 = \vert \langle \psi \vert \psi_n  
\rangle \vert^2$ is interpreted as non-thermal probability distribution related  to the expansion 
$\psi = \sum_{n  \in I_q} \langle \psi_n \vert \psi \rangle \psi_n$. 
Because in principle $supp ( \chi_\psi)$ covers all phase space $M$ the subsets 
where $\chi_\psi <0$ are necessary to ensure that if $\langle \psi \vert
\psi'  \rangle =0$ then   $(\chi_{ \psi}, \chi_{ \psi ' } ) =0$. To illustrate this 
situation in the case $M= T^*\mathbb{R}$ let us consider the orthogonal 
states $\psi_+$ and $\psi_-$, $\psi_\pm = \eta_\pm ( \psi_{G,d} \pm 
\psi_{G, -d})$, where $\psi_{G, \pm d} (x) = (c/ \pi)^{1/4} \re^{
- c (x \mp d)^2/2}$ are Gaussian wave packets localized at $x= \pm d$,
 $c$ is a constant ($c = m \omega / \hbar$ for the harmonic oscillator ground state),
and $\eta_\pm = (2 \pm 2 \re^{-c d^2})^{-1/2}$ are normalization factors. The 
corresponding Wigner functions are $\rho_\pm = \eta_\pm^2 ( \rho_d + \rho_{-d} 
\pm \rho_i)$ where
\begin{equation}
\rho_{ \pm d}  (x,p) = \frac{1}{\pi \hbar} \re^{ - c (x \mp d)^2 - p^2 /c \hbar^2} ~~,~~
 \rho_i (x,p) = \frac{2}{\pi \hbar} \re^{ - c x^2 - p^2 /c \hbar^2}   \cos (2dp/\hbar) ~~,
\end{equation}
such that if $d>0$ then $( \rho_+, \rho_-) =0$. 
The negative values of $\rho_\pm$ indicate that for the localized distribution
$\Delta_{q_0 p_0} (q,p) = \delta(q-q_0) \delta(p-p_0) $, the operator $\hat{\Delta}_{q_0p_0}$,
\begin{equation}
(\hat{\Delta}_{q_0p_0})_{ab} = \frac{2}{\hp} (\hat{\Pi}_{q_0})_{ab} \re^{\ri (a-b) p_0/ \hbar} ~~,~~ 
\hat{\Pi}_{q_0} \psi(x) = \psi(2q_0-x)  ~~,
\end{equation}  
is not positive. For instance, if $(q_0,p_0) =(0,0)$ then $\hat{\Delta}_{00} = 2 \hat{\Pi}_0 / \hp$,
$\hat{\Pi}_0 \psi_\pm = \pm \psi_\pm$, and $\rho_\pm (0,0) = (\rho_\pm , \Delta_{00} ) = \langle 
\psi_\pm \vert \hat{\Delta}_{00} \vert \psi_\pm \rangle = \pm 2 / \hp$. \\ \indent
The quantum cells $q_\psi$ associated to the elements $\chi_\psi \in C_{\hat{H}} $ 
may resemble in particular situations the cells defined by the 
integrality conditions of the old quantum mechanics, but to obtain a more precise description,  in
which $\chi_{\psi_n} \vert_{q_{\psi n'}} = \delta_{nn'}$, a local parametrization of $M$ by 
non-canonical coordinates might be necessary. \\ \indent
For a Wigner function $\rho_\psi$ the quantum entropy defined by $\cs_q (\rho) = -k_B Tr ( \hat{\bar{\rho}} 
\ln \hat{\bar{\rho}} ) $ vanishes, ($\cs_q (\rho_\psi) =0$), while the entropy of (\ref{entr}), 
$\cs (\rho_\psi) = - k_B (\rho_\psi, \ln \bar{\rho}_\psi) = - k_B Tr ( \hat{\bar{\rho}}_\psi \hat{\ln} 
\bar{\rho}_\psi)$, in general is complex.

\end{document}